\documentclass[conference, a4paper]{IEEEtran}
\IEEEoverridecommandlockouts
\usepackage{cite}
\usepackage{amsmath,amssymb,amsfonts}
\usepackage{algorithm,algorithmic}
\usepackage{graphicx}
\usepackage{mathtools}
\usepackage{textcomp}
\usepackage{xcolor}
\usepackage{subcaption}
\usepackage{tabularx,booktabs}
\usepackage{multirow}
\usepackage{caption}
\usepackage[scr=dutchcal,cal=euler]{mathalfa}
\usepackage{tikz}
\usetikzlibrary{arrows,arrows.meta,automata,positioning,shapes.geometric,calc}
\usepackage{pgfplots}
\usepgfplotslibrary{polar}
\usepgflibrary{shapes.geometric}
\pgfplotsset{my style/.append style={axis x line=middle, axis y line=
middle, xlabel={$x$}, ylabel={$y$}, axis equal}}
\def\BibTeX{{\rm B\kern-.05em{\sc i\kern-.025em b}\kern-.08em
    T\kern-.1667em\lower.7ex\hbox{E}\kern-.125emX}}

\tikzstyle{block} = [draw, fill=blue!20, rectangle, 
    minimum height=1cm, minimum width=1.5cm]
\tikzstyle{sum} = [draw, fill=black!20, circle, node distance=1.5cm]
\tikzstyle{input} = [coordinate]
\tikzstyle{output} = [coordinate]
\tikzstyle{pinstyle} = [pin edge={to-,thin,black}]

\begin{document}

\title{Fault-Tolerant Control Design in Scrubber Plant with Fault on Sensor Sensitivity\\
}

\author{\IEEEauthorblockN{\textsuperscript{1}Moh. Kamalul Wafi, \textsuperscript{2}Katherin Indriawati}
\IEEEauthorblockA{\textsuperscript{1,2}\textit{Department of Engineering Physics, Institut Teknologi Sepuluh Nopember, Indonesia}}
\IEEEauthorblockA{\textsuperscript{1}kamalul.wafi@its.ac.id, \textsuperscript{2}katherin@ep.its.ac.id}
}

\maketitle

\begin{abstract}
The concept of fault-tolerant control has extensively been explored with various mapping of development. It starts from the system characteristic, the robustness of the controller, estimation methods and optimization, to the combination of the faults such that it can touch the true observed system. The mathematical concepts of the scrubber plant taking into account the pressure parameter along with sensing element and actuator are proposed. The data to construct the designs derive from the true values in one of Indonesian company. The performances coming from the simulations depict that the open- and closed-loop system could be the same as those of the real results. Furthermore, the observer is proposed to give the estimates of the states of $(\hat{x})$ and $(\hat{f}_s)$ showing the positive trace on the set-point of the residual fault followed by designing the fault-tolerant control with sensor fault on sensitivity. The scenarios are to give the lack of reading in sensor with $70\%$ and $85\%$ sensitivity and those are contrasted to the system without FTC (only PI controller). The yields portray that the system with FTC could deal with those sensor fault scenarios while its counterpart cannot drawing the faulty performance instead of tracking the set-point. The next project associated with this paper is also mentioned in the last section.
\end{abstract}


\section{Introduction}
The early design of proposing fault-tolerant control (FTC) in the late 20$^{\textrm{th}}$ century was triggered by the failure of switching in telephone networks being aggravated with aircraft crashed \cite{R1}. Moreover, fault diagnostics scheme in non-linear stochastic system with sensor fault along with PID-feedback and Monte-Carlo appeared in just a decade after \cite{R2}. The importance for self-tracking to the faults was discussed in \cite{R3} and \cite{R4}, with additional of switching scenario in the second compared to the first. Beyond that, the distribution of the FTC with certain adaptive method, robust Pareto estimation and various estimation methods compared were suggested in \cite{R5}, \cite{R6}, and \cite{R7} in turn, making this research interest becoming more dynamic and extensive to the industrial system. The scenario of observer and the condition of two faults both on sensor and actuator was written in \cite{R8} whereas the same condition with a tolerance measure and $H_{\infty}-$based observer were carried out by \cite{R9} and \cite{R10}, comprising three patterns of faulty states, either sensor or actuator and their combination. As for the paper, the design of two faults would be in the next project along with scenario of unobservability on the system. In terms of application, many objects with discrete-time design are suitable for this research area as mentioned in \cite{R11}.

Having said the development and advanced scenario with respect to fault-tolerant control, the general overview has been well-written in \cite{R12} saying the comprehensive explanation on it. As regards the paper, the mathematical and concept design of the plant was proposed in \cite{R13} such that it can be used to approach the true system. The theoretical and mathematical design of the system being implemented constitute from \cite{R14} and \cite{R15} respectively. As mentioned in the preceding paragraph regarding the next research, the concept would be aided by the following research to deal with unobservability \cite{R16}.

\section{Proposed Design}
This section enlightens the mathematical models of control system focusing on pressure parameter being applied in the Scrubber. Those analytical models comprise the plant itself, the transmitter and the control valve. Furthermore, the controller, the estimation design along with the reconfigurable control in the fault-tolerant control are also presented as the whole integrated system in this research.

\subsection{Mathematical Model of Plant}
The scrubber (V-100) plant in this research is according to real condition in one of the oil and gas Indonesian companies due to the existing of sensor fault. There are two sort of choices, the wet- and dry-scrubber. As regards the first, the direct contact between the polluted gas and the scrubbing liquid is situated by both pool (absorbing) and sprayer (dissolving) of liquid. Dry-scrubber also being used in this paper as shown in Fig. 1 is to spray granulate absorbent such that it vanishes the sulfur dioxide. The mathematical model is initiated by formulating the mass balance \cite{R12}-\cite{R15} such that,
\begin{figure*}[t!]
	\centering
	\begin{subfigure}[t]{0.32\linewidth}
		\centering
		\includegraphics[width=\textwidth, height=6cm]{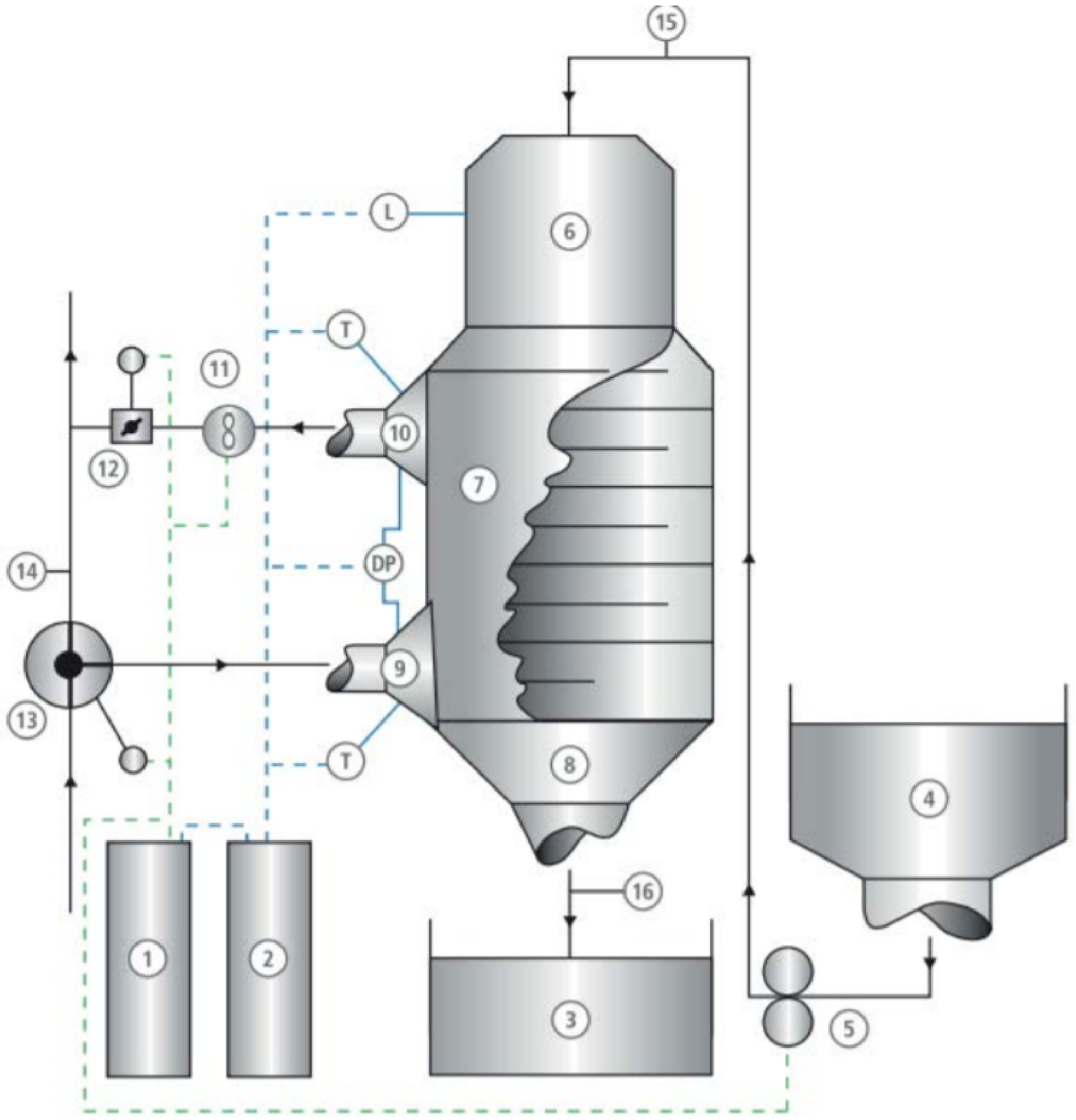}
		\caption{}
		\label{Fig 1a}
	\end{subfigure}
	~
    \begin{subfigure}[t]{0.32\linewidth}
		\centering
		\includegraphics[width=\textwidth, height=6cm]{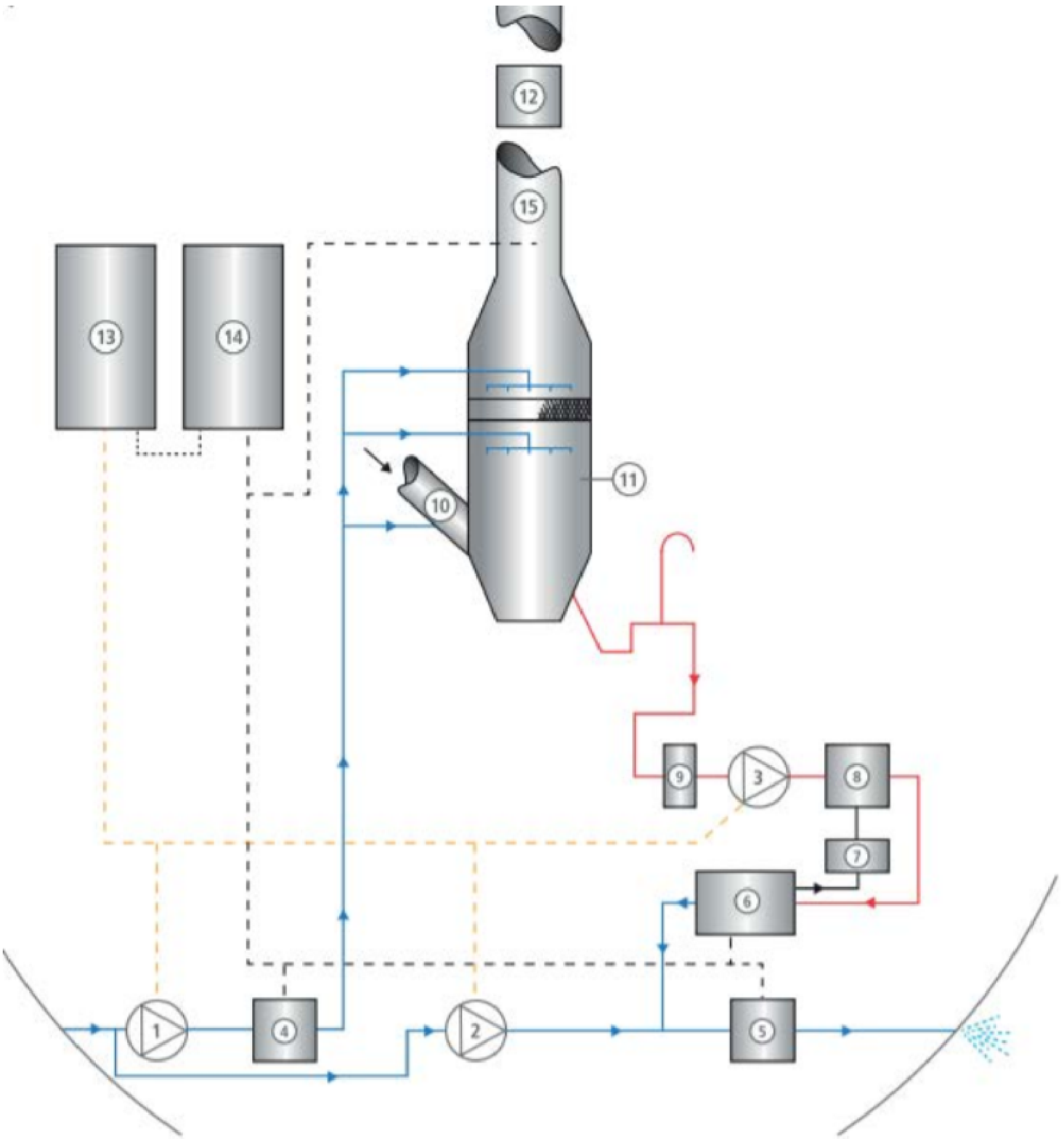}
		\caption{}
		\label{Fig 1b}
	\end{subfigure}
	~
    \begin{subfigure}[t]{0.32\linewidth}
		\centering
		\includegraphics[width=\textwidth, height=6cm]{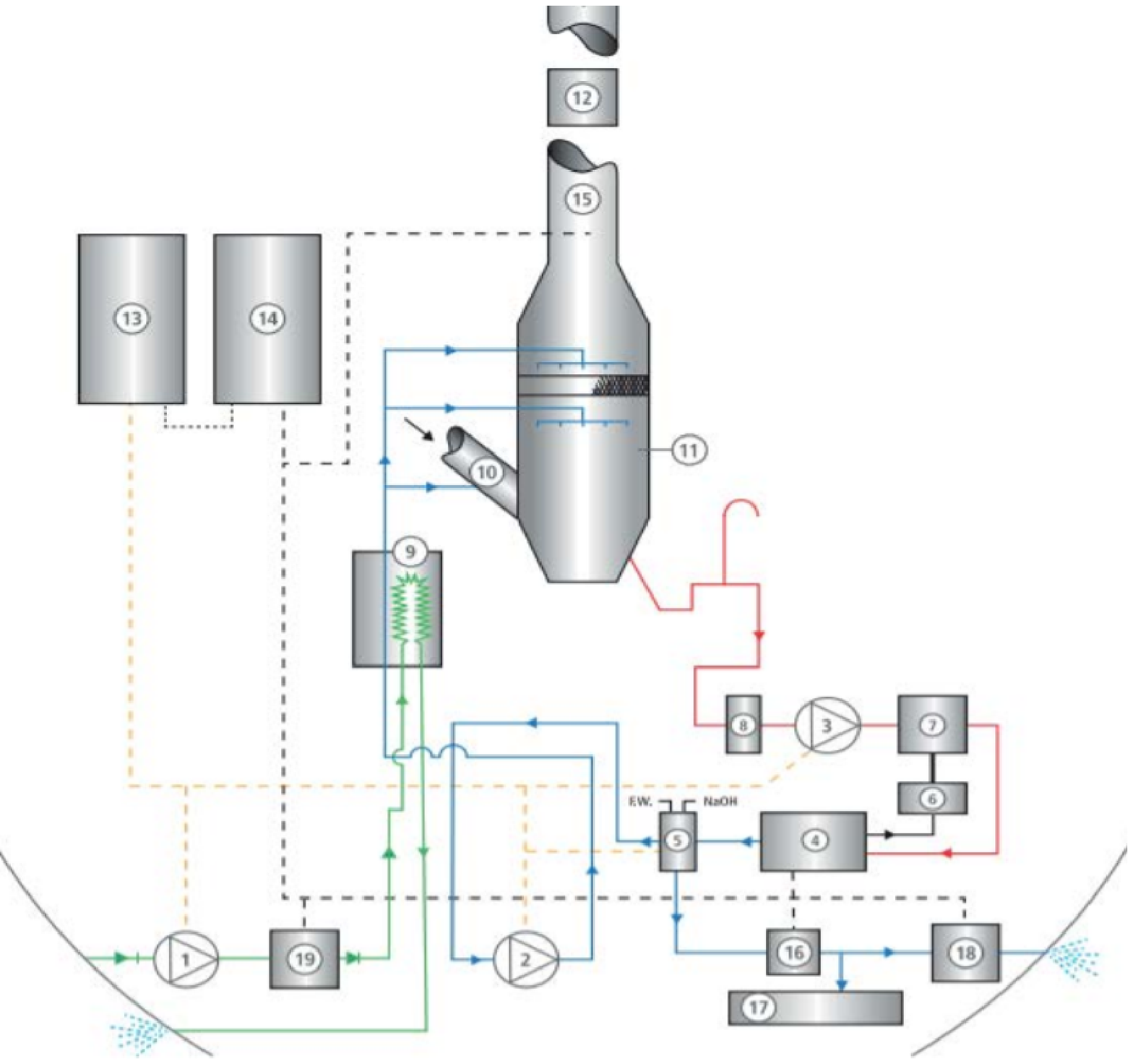}
		\caption{}
		\label{Fig 1c}
	\end{subfigure}
	\label{Figure1}
	\caption{(a). System of dry scrubber; (b). Scrubber open-loop design; (c). Scrubber closed-loop design \cite{R13}}
\end{figure*}
\begin{table*}
    \centering
    \captionsetup{font=small}
    \caption{Description of instruments in Fig. 1 as stated in \cite{R13}}
    \begin{tabular}{c|c|c|c|c}
        \toprule
        No. & Dry Scrubber & Open-loop & Closed-loop & Hybrid\\
        \midrule
        1 & Control cabinet & Salt wash water pump & Cooler sea-water pump & \multirow{19}{*}{\shortstack{Combination of \\ open-loop and \\ closed-loop system}}\\
        2 & Monitoring cabinet & Effluent dilution pump & Wash water supply pump &\\
        3 & Reaction product & Effluent water discharge pump & Effluent water re-circulation pump &\\
        4 & Granulate storage & Salt wash water monitor & Oil \& soot separator &\\
        5 & Monitoring Cabinet & Effluent water monitoring & Alkali (NaOH) Unit &\\
        6 & Granulate silo & Oil \& soot separator & EGC residue tank &\\
        7 & Scrubber reactor & EGC residue tank & Sludge separator &\\
        8 & Reaction product & Sludge separator & Reareation unit &\\
        9 & Exhaust inlet & Reareation unit & Wash water cooler &\\
        10 & Exhaust outlet & Exhaust gas inlet & Exhaust gas inlet &\\
        11 & Exhaust fan & Scrubber & Scrubber &\\
        12 & Isolation valve & Exhaust gas inlet & Exhaust fan &\\
        13 & 3-way inlet/Bypass valve & Control fan & Control cabinet &\\
        14 & Exhaust bypass manifold & Monitoring \& alarm cabinet & Monitoring \& alarm cabinet &\\
        15 & Granulate absorption inlet & Exhaust gas outlet & Exhaust gas outlet &\\
        16 & Reaction product outlet & & Bleed-off treatment unit &\\
        17 & & & Holding tank &\\
        18 & & & Effluent water monitoring &\\
        19 & & & Wash water cooling monitoring &\\
        \bottomrule
    \end{tabular}
    \label{tab:my_label}
\end{table*}
\begin{align}
    \frac{d}{dt}\left(\rho_l V_l + \rho_g V_g\right) = q_{i} - q_{o(l)} - q_{o(g)}
    \label{Eq 1}
\end{align}
and,
\begin{align}
    \frac{d}{dt}\left(\rho_l V_l + \rho_g V_g\right) = V_l \frac{\partial \rho_l}{\partial t} + \rho_l \frac{\partial V_l}{\partial t} + V_g \frac{\partial \rho_g}{\partial t} + \rho_g \frac{\partial V_g}{\partial t}
    \label{Eq 2}
\end{align}
where $q_i, \rho, V$ and $q_o$ are inlet flow (kg/s), density (kg/m$^3$), volume (m$^3$) and outlet flow with respect to gas $(g)$ and liquid $(l)$. The energy balance should also be considered, therefore  
\begin{subequations}
\begin{align}
    E_2 - E_1 &= \left(E_{k_2} - E_{k_1}\right) + \left(E_{p_2} - E_{p_1}\right) + \left(U_2 - U_1\right)\label{Eq 3a}\\
    \Delta E &= \Delta E_k + \Delta E_p + \Delta U \label{Eq 3b}
\end{align}
\end{subequations}
which can be written as Eq. (\ref{Eq 4a}) along with its flow rate (\ref{Eq 4b}). 
\begin{subequations}
\begin{align}
    Q - W &= \Delta E_k + \Delta E_p + \Delta U \label{Eq 4a}\\
    \dot{Q} - \dot{W} &= \frac{d}{dt}\left(\Delta E_k + \Delta E_p + \Delta U\right) \label{Eq 4b}
\end{align}
\end{subequations}
From Eqs. (\ref{Eq 3a}) and (\ref{Eq 3b}) to Eq. (\ref{Eq 1}), the equation is formulated as follows,
\begin{subequations}
\begin{align}
    \frac{d}{dt}E &= E_i - E_o + Q \longrightarrow E = \rho_g h_g V_g + \rho_l h_l V_l\label{Eq 5a}\\
    &= q_i h_i - q_{o(l)} h_{o(l)} - q_{o(g)} h_{o(g)} \label{Eq 5b}
\end{align}
\end{subequations}
and the equations for each liquid and gas inside the scrubber are given below taking into account the inlet $(\dot{m}_i)$ and outlet $(\dot{m}_o)$ flow along with their enthalpy (input-output) $h_i$ (J/kg) and $h_o$ (J/kg), 
\begin{subequations}
\begin{align}
    \frac{d}{dt}\left(\rho_l h_l V_l\right) &= \rho_{i(l)} h_{i(l)} \dot{m}_{i(l)} - \rho_{o(l)} h_{o(l)} \dot{m}_{o(l)}\label{Eq 6a}\\
    \frac{d}{dt}\left(\rho_g h_g V_g\right) &= \rho_{i(g)} h_{i(g)} \dot{m}_{i(g)} - \rho_{o(g)} h_{o(g)} \dot{m}_{o(g)}\label{Eq 6b}
\end{align}
\end{subequations}
Furthermore, pressure balance is also considered as the sensing element is pressure itself such that the mathematical theory is proposed in the following,
\begin{subequations}
\begin{align}
    \frac{d}{dt}p &= \frac{\rho h g}{A}\left(\dot{m}_i - \dot{m}_o\right)\label{Eq 7a}\\
    &= \frac{\rho_i h_i g_i \dot{m}_i}{A} - \frac{\rho_{o(g)} h_{o(g)} g_{o(g)} \dot{m}_{o(g)}}{A}\label{Eq 7b}
\end{align}
\end{subequations}
where $p$ is pressure (Psi) with $\dot{m}_{o(g)}$ equals to $k\sqrt{p}$ which should be linearized applying Taylor series by avoiding the higher-order of the equation Eq. (\ref{Eq 9}) resulting Eq. (\ref{Eq 8}).
\begin{align}
    k\sqrt{p} = k\sqrt{p_o} + \frac{k}{2\sqrt{p_o}} (p - p_o) \label{Eq 8}
\end{align}
By substituting the result above, Eq. (\ref{Eq 7b}) is then transformed to Eq. (\ref{Eq 11}) after being altered into Laplace transform and some algebra. The transfer function of scrubber is described in Eq. (\ref{Eq 12}) with $p_o, K_s$ and $\tau_s$ are pressure output, the gain and the time constant of scrubber. Those parameters are obtained from the measurement with control system as depicted in Fig. (\ref{Figure2})
\begin{figure*}[t!]
\begin{gather}
    k\sqrt{p} = k\sqrt{p_o} + \left.\frac{d\left(k\sqrt{p}\right)}{dp}\right|_{p = p_o} (p - p_o) + \left.\frac{d^2\left(k\sqrt{p}\right)}{dp^2}\right|_{p = p_o} \frac{(p - p_o)^2}{2!} + \left.\frac{d^3\left(k\sqrt{p}\right)}{dp^3}\right|_{p = p_o} \frac{(p - p_o)^3}{3!} + \cdots\label{Eq 9}\\[1pt]
    \frac{dp}{dt} + \frac{\rho_{o(g)} h_{o(g)} g}{A}\left(k\sqrt{p_o} + \frac{k}{2\sqrt{p_o}} (p - p_o)\right) = \frac{\rho_i h_i g \dot{m}_i}{A}\label{Eq 10}\\[1pt]
    P(s)\left(\frac{2A\sqrt{p_o}}{\rho_{o(g)} h_{o(g)} g k}s + 1 \right) = \left(\frac{2\sqrt{p_o}}{k}\right)\frac{\rho_i h_i g}{\rho_{o(g)} h_{o(g)} g} \dot{M}_i(s) \longrightarrow \textrm{after Laplace and divided by } \frac{\rho_{o(g)} h_{o(g)} g}{A}\left(\frac{k}{2\sqrt{p_o}}\right)\label{Eq 11}\\[1pt]
    \frac{P(s)}{\dot{M}_i(s)} = \dfrac{\left(\dfrac{2\sqrt{p_o}}{k}\right)\dfrac{\rho_i h_i g}{\rho_{o(g)} h_{o(g)} g}}{\dfrac{2A\sqrt{p_o}}{\rho_{o(g)} h_{o(g)} g k}s + 1} \coloneqq \frac{K_s}{\tau_s s + 1}\label{Eq 12}
\end{gather}
\end{figure*}
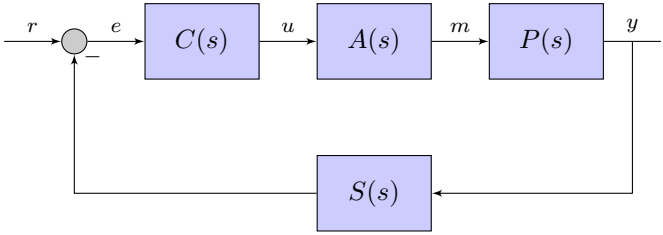
\begin{figure}[!t]
    \captionsetup{font=small}
    \centering
    \begin{tikzpicture}[auto,>=latex']
    \node [coordinate] (C1) {};
    \node [sum, right = .75cm of C1] (Sum) {};
    \node [block, right = .75cm of Sum] (Cont) {$C(s)$};
    \node [block, right = .75cm of Cont] (Act) {$A(s)$};
    \node [block, below = 1cm of Act] (Sens) {$S(s)$};
    \node [block, right = .75cm of Act] (Plant) {$P(s)$};
    \node [coordinate, right = 0.375cm of Plant] (C2) {};
    \node [coordinate, right = 0.375cm of C2] (C3) {};
    
    \draw [draw, ->] (C1) -- node{\footnotesize $r$} (Sum);
    \draw [draw, ->] (Sum) -- node{\footnotesize $e$} (Cont);
    \draw [draw, ->] (Cont) -- node{\footnotesize $u$} (Act);
    \draw [draw, ->] (Act) -- node{\footnotesize $m$} (Plant);
    \draw [draw, -] (Plant) -- node{\footnotesize $y$} (C3);
    \draw [draw, ->] (C2) |- (Sens);
    \draw [draw, ->] (Sens) -| node[pos=0.99, right] {\footnotesize $-$} (Sum);
    
    \end{tikzpicture}
    \caption{A loop of control system with controller ($C$), actuator ($A$), plant ($P$) and sensing element ($S$)}
    \label{Figure2}
\end{figure}

\subsection{Mathematical Model of Actuator and Sensing}
Pressure transmitter (PT-0105) with diaphragm-based is implemented so that what is measure will be transferred into $4-20$ mA (DC). Its mathematical design can be approached as given below,
\begin{align}
    G_t(s) = \frac{K_t}{\tau_t s + 1} \longrightarrow K_t = \frac{O_{\textrm{max}} - O_{\textrm{min}}}{I_{\textrm{max}} - I_{\textrm{min}}} \label{Eq 13}
\end{align}
where $K_t$ is gain transmitter with the span comparison from output over input while $\tau_t$ is the time constant of the transmitter. With respect to actautor being used, it is PV-0105 to maintain the flow as the set point. The control valve gain $K_v$ (mmscfd/mA) is constructed from $\mathcal{G}_v$ (mmscfd/psi) and $\mathcal{G}_{\textrm{I/P}}$ converting $4-20$ (mA) to pneumatic signal $3-15$ (psi) which can be seen in Eq. (\ref{Eq 14})
\begin{align}
    \mathcal{G}_v = \frac{\textrm{span output}}{\textrm{span input}} \textrm{ and } \mathcal{G}_{\textrm{I/P}} = \frac{\textrm{span output}}{\textrm{span input}} \label{Eq 14}
\end{align}
such that,
\begin{align}
    K_v = \mathcal{G}_v \cdot \mathcal{G}_{\textrm{I/P}} \label{Eq 15}
\end{align}
The time constant of control valve is then generated in Eq. (\ref{Eq 16}). $T_v$ is full stroking time with $\frac{Y_c}{C_v}$, $Y_c$ is the factor stroking time and $R_v$ is the constant between inherent-time over stroke, constituting either $0.03$ or $0.03$ for diaphragm- or piston-based actuator while $\Delta V$ is the change fraction, such that
\begin{gather}
    \tau_v = T_v(\Delta V + R_v) \longrightarrow \Delta V = \frac{q_{\textrm{max}} - q_{\textrm{min}}}{q_{\textrm{max}}} \label{Eq 16}\\
    G_v = \frac{K_v}{\tau_v s + 1} \coloneqq \frac{\dot{M}_i(s)}{U(s)} \label{Eq 17}
\end{gather}
The transfer function of actuator is then written as Eq. (\ref{Eq 17}) which equals to the manipulated variable of $\dot{M}_i$ (Laplace) over signal input $U$.
\begin{figure}[!t]
    \captionsetup{font=small}
    \centering
    \begin{tikzpicture}[auto,>=latex']
    \node [coordinate] (C1) {};
    \node [sum, right = .75cm of C1] (Sum) {};
    \node [coordinate, right = .75cm of Sum] (C2) {};
    \node [block, right = .5cm of C2] (Int) {$\displaystyle K_i \int_0^t e(t')\,dt'$};
    \node [block, above = .5cm of Int] (Prop) {$K_p e(t)$};
    \node [block, below = .5cm of Int] (Deriv) {$\displaystyle K_d \frac{de(t)}{dt}$};
    \node [sum, right = .5cm of Int] (Sum2) {};
    \node [block, right = 0.75cm of Sum2] (Plant) {Plant};
    \node [coordinate, right = 0.375cm of Plant] (C3) {};
    \node [coordinate, right = 0.375cm of C3] (C4) {};
    \node [coordinate, below = 2.5cm of C3] (C5) {};
 
    \draw [draw, ->] (C1) -- node[above] {\footnotesize $r$} (Sum);
    \draw [draw, -] (Sum) -- node {\footnotesize $e$} (C2);
    \draw [draw, ->] (C2) -- (Int);
    \draw [draw, ->] (C2) |- (Prop);
    \draw [draw, ->] (C2) |- (Deriv);
    \draw [draw, ->] (Int) -- (Sum2);
    \draw [draw, ->] (Prop) -| (Sum2);
    \draw [draw, ->] (Deriv) -| (Sum2);
    \draw [draw, ->] (Sum2) -- node[above] {\footnotesize $u$} (Plant);
    \draw [draw, ->] (Plant) -- node[above] {\footnotesize $y$} (C4);
    \draw [draw, -] (C3) -- (C5);
    \draw [draw, ->] (C5) -| node[pos=0.99, right] {\footnotesize $-$} (Sum);
    
    \draw [color=blue,dashed](2,-2.25) rectangle (5,2.25);
    \end{tikzpicture}
    \caption{Scheme of Disturbance Observer $\mathcal{O}$}
    \label{Figure3}
\end{figure}
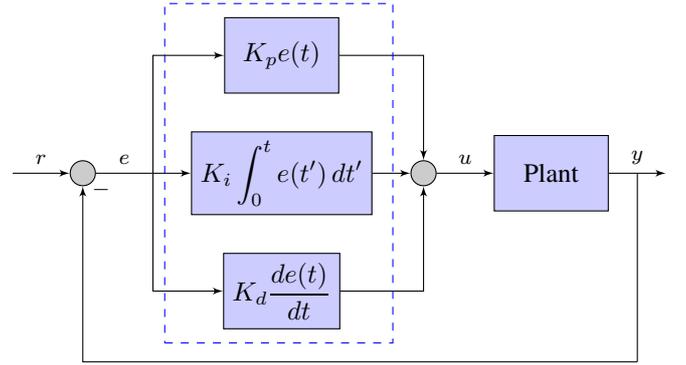

\subsection{Mathematical Model of Controller}
\begin{figure}[t!]
    \captionsetup{font=small}
    \centering
    \includegraphics[width=0.9\linewidth]{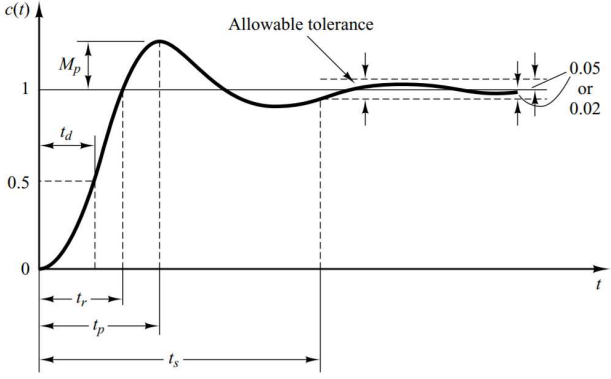}
    \caption{Analysis of transient response \cite{R15}}
    \label{Figure 4}
\end{figure}

The transient response is the key to yield the best performance so that it should be inside the good-threshold. There are four major analysis presented in the following equations, where $\omega_d = \omega_n\sqrt{1 - \zeta^2}$ and $\sigma = \zeta\omega_n$. $t_r, t_p, t_s$ refer to rise-time, peak-time and settling-time along with maximum overshoot $M_p$ which are clearly depicted in Fig. (\ref{Figure 4}), therefore
\begin{subequations}
\begin{align}
    t_r &= \frac{\pi - \beta}{\omega_d}(s) \longrightarrow \beta = \tan^{-1}\frac{\omega_d}{\sigma}(\textrm{rad})\label{Eq 18a}\\
    t_p &= \frac{\pi}{\omega_d}(s)\label{Eq 18b}\\
    t_s &= \frac{4}{\sigma}(s)\rightarrow 2\% \quad t_s = \frac{3}{\sigma}(s)\rightarrow 5\%\label{Eq 18c}\\
    M_P &= e^{-\left(\frac{\sigma}{\omega_d}\right)\pi} (\%)\label{Eq 18d}
\end{align}
\end{subequations}
The controller being implemented in this system is \textbf{PI} with scenario as highlighted in Eq. (\ref{Eq 19}), detailing $K_p$, $K_i = \frac{K_p}{T_i}$ and $K_d = K_pT_d$. The success of the performance is determined by generating the best fit for these gain 
\begin{align}
    u(t) = K_p\left[e(t) + \frac{1}{T_i}\int_0^{t'} e(t')\, dt' + T_d\frac{de(t)}{dt}\right] \label{Eq 19}
\end{align}

\subsection{Mathematical Model of Observer}
The basic dynamical control system is illustrated in these two equations, 
\begin{subequations}
\begin{align}
    \dot{x} &= Ax + Bu\label{Eq 20a}\\
    y &= Cx + \mathscr{F}f_s \label{Eq 20b}
\end{align}
\end{subequations}
The matrices of $A, B, C, \mathscr{F}$ is for the system, input, output and the fault affected by two parameters, pressure and the inlet flow acting as the state of the system $x$. This matrix is built from two transfer functions of both plant (scrubber) and actuator (PV-0105) as mentioned in Eqs. (\ref{Eq 12}) and (\ref{Eq 17}) followed by inverse-Laplace, such that
\begin{subequations}
\begin{align}
    P(s)\left(\tau_s s + 1\right) &= K_s \dot{M}_i(s)\label{Eq 21a}\\
    sP(s) &= \frac{-P(s) + K_s\dot{M}_i(s)}{\tau_s}\label{Eq 21b}\\
    \frac{dp(t)}{dt} &= \frac{-p(t) + K_s\dot{m}_i(t)}{\tau_s} \longrightarrow \textrm{inverse} \label{Eq 21c}
\end{align}
\end{subequations}
and,
\begin{subequations}
\begin{align}
    \dot{M}_i(s)\left(\tau_v s + 1\right) &= K_v U(s)\label{Eq 22a}\\
    s\dot{M}_i(s) &= \frac{-\dot{M}_i(s) + K_vU(s)}{\tau_v}\label{Eq 22b}\\
    \frac{d\dot{m}_i(t)}{dt} &= \frac{-\dot{m}_i(t) + K_v u(t)}{\tau_v} \longrightarrow \textrm{inverse} \label{Eq 22c}
\end{align}
\end{subequations}
After computing these two formulas, the whole state-space can be arranged as given in Eqs. (\ref{Eq 23a}) and (\ref{Eq 23b}),
\begin{subequations}
\begin{align}
    \frac{d}{dt}\begin{bmatrix}
    p(t)\\
    \dot{m}_i\end{bmatrix} &= \begin{bmatrix}
    -\dfrac{1}{\tau_s} & \dfrac{K_s}{\tau_s}\\[7pt]
    0 & -\dfrac{1}{\tau_v}\end{bmatrix}\begin{bmatrix}
    p(t)\\
    \dot{m}_i\end{bmatrix} + \begin{bmatrix}
    0 \\[7pt]
    \dfrac{K_v}{\tau_v}\end{bmatrix} u(t)\label{Eq 23a}\\
    y &= \begin{bmatrix}
    1 & 0\\
    0 & 1\end{bmatrix}\begin{bmatrix}
    p(t)\\
    \dot{m}_i\end{bmatrix} + \begin{bmatrix}
    1\\
    0\end{bmatrix}f_s \label{Eq 23b}
\end{align}
\end{subequations}
To compensate the sensor fault as design in this paper, it is required to modify the state-space by inserting a new variable of $\dot{\xi}$ as in Eq. (\ref{Eq 24}) to the original states resulting the Eq. (\ref{Eq 25a}) and (\ref{Eq 25b}), therefore,
\begin{align}
    \dot{\xi} &= \Phi(y - \xi)\nonumber\\
    &= \Phi Cx + \Phi\mathscr{F}f_s - \Phi\xi \label{Eq 24}
\end{align}
such that,
\begin{subequations}
\begin{align}
    \begin{bmatrix}
    \dot{x}\\
    \dot{\xi}\end{bmatrix} &= \begin{bmatrix}
    A & \textbf{O}\\
    \Phi C & -\Phi\end{bmatrix}\begin{bmatrix}
    x\\
    \xi\end{bmatrix} + \begin{bmatrix}
    B\\
    \textbf{O}\end{bmatrix}u + \begin{bmatrix}
    \textbf{O}\\
    \xi\mathscr{F}\end{bmatrix}f_s \label{Eq 25a}\\
    y_e &= \begin{bmatrix}
    \textbf{O} & I\end{bmatrix}\begin{bmatrix}
    x\\
    \xi\end{bmatrix} \label{Eq 25b}
\end{align}
\end{subequations}
Those two modified equations above can be simplified into these two equations Eqs. (\ref{Eq 26a}) and (\ref{Eq 26b}) respectively, yielding
\begin{subequations}
\begin{align}
    \dot{x}_e &= A_e x_e + B_e u + \mathscr{F}_e f_s \label{Eq 26a}\\
    y_e &= C_e x_e \label{Eq 26b}
\end{align}
\end{subequations}
These formulas are applied in the observer in estimating the state ($\hat{x}_e$), fault ($\hat{f}_s$) and output ($\hat{y}_e$) as highlighted in Eqs. (\ref{Eq 27a}), (\ref{Eq 27b}), and (\ref{Eq 27c}), constituting
\begin{subequations}
\begin{align}
    \dot{\hat{x}}_e &= A_e \hat{x}_e + B_e u + \mathscr{F}_e \hat{f}_s + L_x(y_e - \hat{y}_e)\label{Eq 27a}\\
    \dot{\hat{f}}_s &= L_f(y_e - \hat{y}_e)\label{Eq 27b}\\
    \hat{y}_e &= C_e \hat{x}_e \label{Eq 27c}
\end{align}
\end{subequations}
where the forms of matrices are then written in Eqs. (\ref{Eq 28a}) and (\ref{Eq 28b}), 
\begin{subequations}
\begin{align}
    \begin{bmatrix}
    \dot{\hat{x}}_e\\
    \dot{\hat{f}}_s\end{bmatrix} &= \begin{bmatrix}
    A_e & \mathscr{F}_e\\
    \textbf{O} & \textbf{O}\end{bmatrix}\begin{bmatrix}
    \hat{x}_e\\
    \hat{f}_s\end{bmatrix} + \begin{bmatrix}
    B_e\\
    \textbf{O}\end{bmatrix}u + \begin{bmatrix}
    L_x\\
    L_f\end{bmatrix}(y_e - \hat{y}_e)\label{Eq 28a}\\
    \hat{y}_e &= \begin{bmatrix}
    C_e & \textbf{O}\end{bmatrix}\begin{bmatrix}
    \hat{x}_e\\
    \hat{f}_s\end{bmatrix} \label{Eq 28b}
\end{align}
\end{subequations}
along with its simplification equations as Eqs. (\ref{Eq 29a}) and (\ref{Eq 29b}),
\begin{subequations}
\begin{align}
    \dot{x}_\gamma &= A_\gamma x_\gamma + B_\gamma u + K_\gamma f_s\label{Eq 29a}\\
    \hat{y}_e &= C_\gamma x_\gamma \label{Eq 29b}
\end{align}
\end{subequations}

\subsection{Fault-Tolerant Control Design}
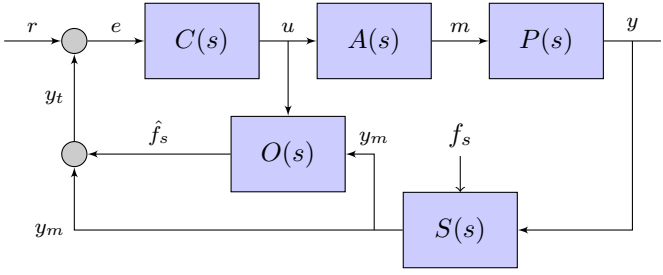
\begin{figure}[!t]
    \captionsetup{font=small}
    \centering
    \begin{tikzpicture}[auto,>=latex']
    \node [coordinate] (C1) {};
    \node [sum, right = .75cm of C1] (Sum) {};
    \node [sum, below of = Sum, node distance=1.5cm] (Sum2) {};
    \node [block, right = .75cm of Sum] (Cont) {$C(s)$};
    \node [coordinate, right = .375 of Cont] (C4) {};
    \node [block, below of= C4, node distance=1.5cm] (Obs) {$O(s)$};
    \node [block, right = .75cm of Cont] (Act) {$A(s)$};
    \node [coordinate, right = .375 of Act] (C5) {};
    \node [block, below = 2cm of C5, pin={[pinstyle]above:$f_s$}] (Sens) {$S(s)$};
    \node [coordinate, left = .375 of Sens] (C6) {};
    \node [block, right = .75cm of Act] (Plant) {$P(s)$};
    \node [coordinate, right = 0.375cm of Plant] (C2) {};
    \node [coordinate, right = 0.375cm of C2] (C3) {};
    
    \draw [draw, ->] (C1) -- node{\footnotesize $r$} (Sum);
    \draw [draw, ->] (Sum) -- node{\footnotesize $e$} (Cont);
    \draw [draw, ->] (Cont) -- node{\footnotesize $u$} (Act);
    \draw [draw, ->] (Act) -- node{\footnotesize $m$} (Plant);
    \draw [draw, -] (Plant) -- node{\footnotesize $y$} (C3);
    \draw [draw, ->] (C2) |- (Sens);
    \draw [draw, ->] (C6) |- node[above] {\footnotesize $y_m$} (Obs);
    \draw [draw, ->] (C4) -| (Obs);
    \draw [draw, ->] (Sens) -| node {\footnotesize $y_m$} (Sum2);
    \draw [draw, ->] (Obs) -- node[above] {\footnotesize $\hat{f}_s$} (Sum2);
    \draw [draw, ->] (Sum2) -- node {\footnotesize $y_t$} (Sum);
    
    \end{tikzpicture}
    \caption{Scheme of Disturbance Observer $O$}
    \label{Figure5}
\end{figure}

The fault-tolerant control (FTC) comprising two most renowned designs, passive-FTC and active-FTC, guarantees the system to accommodate the failures. The high performance along with its reliability are the aims to formulate this mechanism. The robust fixed-controller with apriori faults accounts for the passive-FTC while its counterpart is supposed to be more dynamic to the faulty states such that it maintains throughout the stability system. With respect to the scheme, fault detection diagnosis (FDD) along with reconfigurable control exist in the active-FTC whereas the passive-FTC does not require those two\cite{R12}. From Fig. (\ref{Figure5}), it can be written that,
\begin{align}
    y_m &= y + f_s \\
    y_t &= y_m - \hat{f}_s
\end{align}
and,
\begin{align}
    e = r - y_t
\end{align}
where $r$, $e$, and $u$ are the reference signal, the error, and the control signal in turn. The $y_m$ is the measured variable which is the addition of the true value and the faulty state $f_s$. $y_t$ as the estimated output would be returned compared to the reference.

\section{Simulation Results}
\subsection{Mathematical Model}
The final section discusses the success of the proposed mathematical designs. The volume ($V$), diameter ($d$), height ($H$) and the pressure set of the scrubber are $2.5$ m$^3$, $1.07$ m, $2.4$ m and $24$ bar. The specific gravity of the liquid ($\gamma_l$) and gas ($\gamma_g$) are $0.726$ and $1.173$. The density of input ($\rho_i$) and output ($\rho_o$) flow along with their enthalpy ($h_i$) and ($h_o$) are $5.2$ kg/m$^3$, $4.9$ kg/m$^3$, $4.9$ J/kg, and $4.1$ J/kg in turn. $k$ and $p_o$ equal to $1$ and $348.091$. In terms of sensing element and actuator, $G_t$ is $1$ and $\mathcal{G}_{\textrm{I/P}}$ accounts for $3-15$ psi over $4-20$ mA while the span of gas makes up $12-16$ mmscfd. $Y_c$ and $C_v$ are $0.68$ and $117$ respectively. As for controller, $K_p$ and $T_i$ are $0.1396$ and $0.3294$
\begin{align}
    A_{\gamma} &= \left[\begin{array}{c c c c c}
	-5.0250 & 277.4500 & 0   & 0  & 0 \\
	0       & -3.9680  & 0   & 0  & 0 \\
	1       & 0        & -1  & 0  & 1 \\
	0       & 1        & 0   & -1 & 0 \\
	0       & 0        & 0   & 0  & 0 \end{array}\right],\\
	B_{\gamma} &= \left[\begin{array}{c}
	0 \\
	0.9920 \\
	0 \\
	0 \\
	0 \end{array}\right],\\
	C_{\gamma} &= \left[\begin{array}{r r r r r}
	0 & 0 & 1 & 0 & 0\\
	0 & 0 & 0 & 1 & 0\end{array}\right]
\end{align}
The next is to check pole placement so that the system is inside unit circle and it result in the following,
\begin{align*}
    \delta_1 &= -54.4047 + 33.5101i \quad \delta_4 = -0.1951 \\
    \delta_2 &= -54.4047 - 33.5101i \quad \delta_5 = -0.5291 \\
    \delta_3 &= -2.7588
\end{align*}
where,
\begin{align*}
    K_{\gamma} = \left[\begin{array}{rrrrr}
    -80.0016 &  0.6563 & 8.4656 & -0.2843 & 0.2234\\
     18.9379 & -0.2651 & 3.4306 &  0.0377 & 0.0301\end{array}\right]
\end{align*}
\begin{table}[h!]
    \centering
    \captionsetup{font=small}
    \caption{Fluid composition entering the scrubber V-100}
    \begin{tabular}{c|c|c||c|c|c}
        \toprule
        Gas & Unit & Input & Gas & Unit & Input\\
        \midrule
        Methane & \multirow{8}{*}{CMP} & 0.7331 & n-Heptane & \multirow{8}{*}{CMP} & 0.0052\\
        Ethane & & 0.0631 & n-Octane & & 0.0048\\
        Propane & & 0.0396 & n-Nonane & & 0.0021\\
        i-Butane & & 0.0087 & n-Decane & & 0.0011\\
        n-Butane & & 0.0068 & n-C11 & & 0.0004\\
        i-Pentane & & 0.0038 & n-C12 & & 0.0003\\
        n-Pentane & & 0.0033 & CO$_2$ & & 0.0127\\
        n-Hexane & & 0.0042 & H$_2$O & & 0.1105\\
        \bottomrule
    \end{tabular}
    \label{Table2}
\end{table}
\begin{table}[h!]
    \centering
    \captionsetup{font=small}
    \caption{Design of open-loop compared to the real values}
    \begin{tabular}{c|c|c}
        \toprule
        Real & Simulation & Error $(\%)$\\
        \midrule
        346.113 & 349.6 & 1.01\\
        347.235 & 351   & 1.08\\
        348.091 & 352.1 & 1.15\\
        346.702 & 350.2 & 1.01\\
        344.805 & 347.9 & 0.90\\
        345.921 & 349.9 & 1.15\\
        347     & 350.7 & 1.07\\
        345.065 & 349   & 1.14\\
        342.186 & 345.1 & 0.85\\
        344.237 & 347.7 & 1.01\\
        \bottomrule
    \end{tabular}
    \label{Table3}
\end{table}

\subsection{Numerical Analysis}
\begin{figure*}[t!]
	\centering
	\begin{subfigure}[t]{0.425\linewidth}
		\centering
		\includegraphics[width=\textwidth]{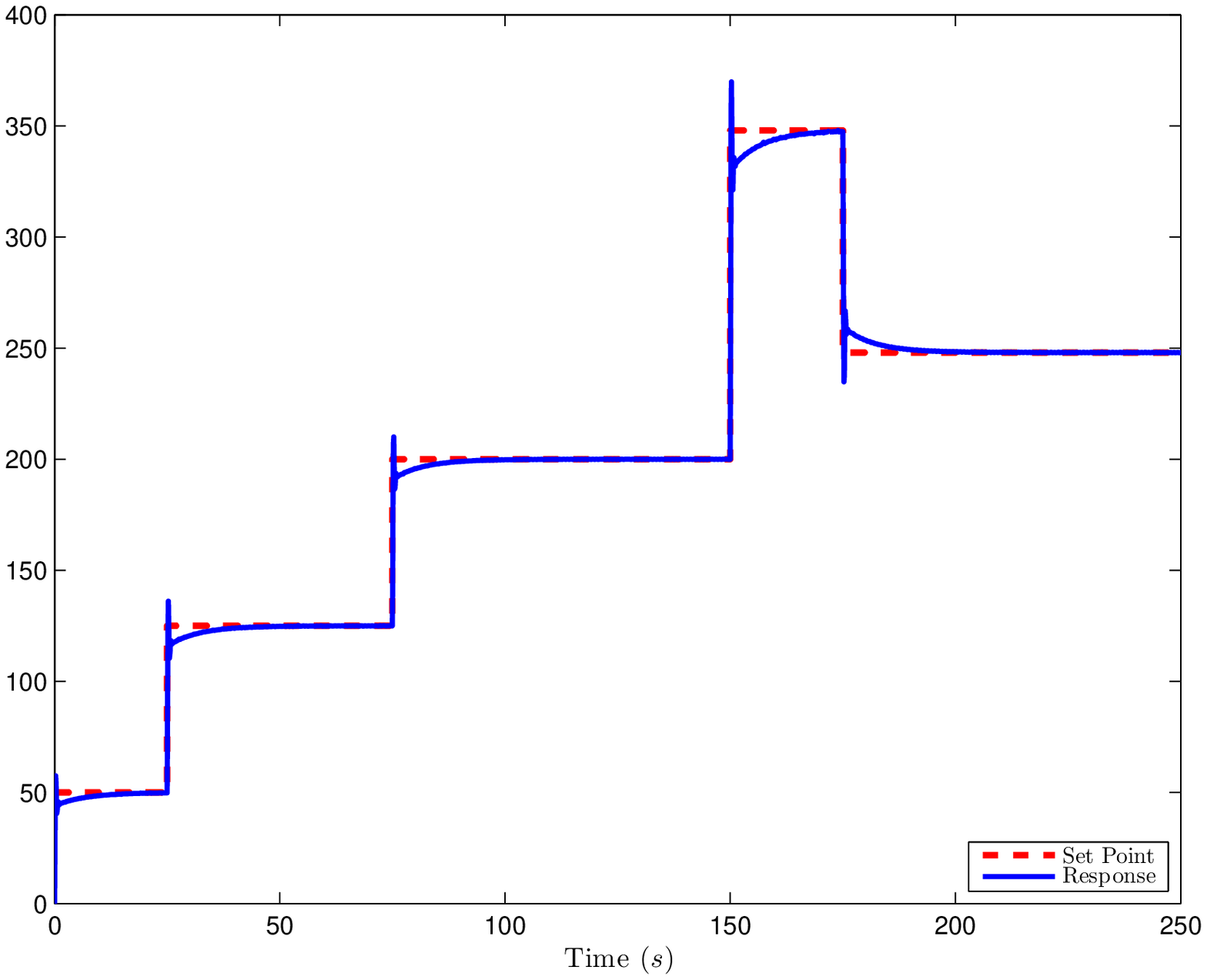}
		\caption{}
		\label{Fig 6a}
	\end{subfigure}
	~
    \begin{subfigure}[t]{0.425\linewidth}
		\centering
		\includegraphics[width=\textwidth]{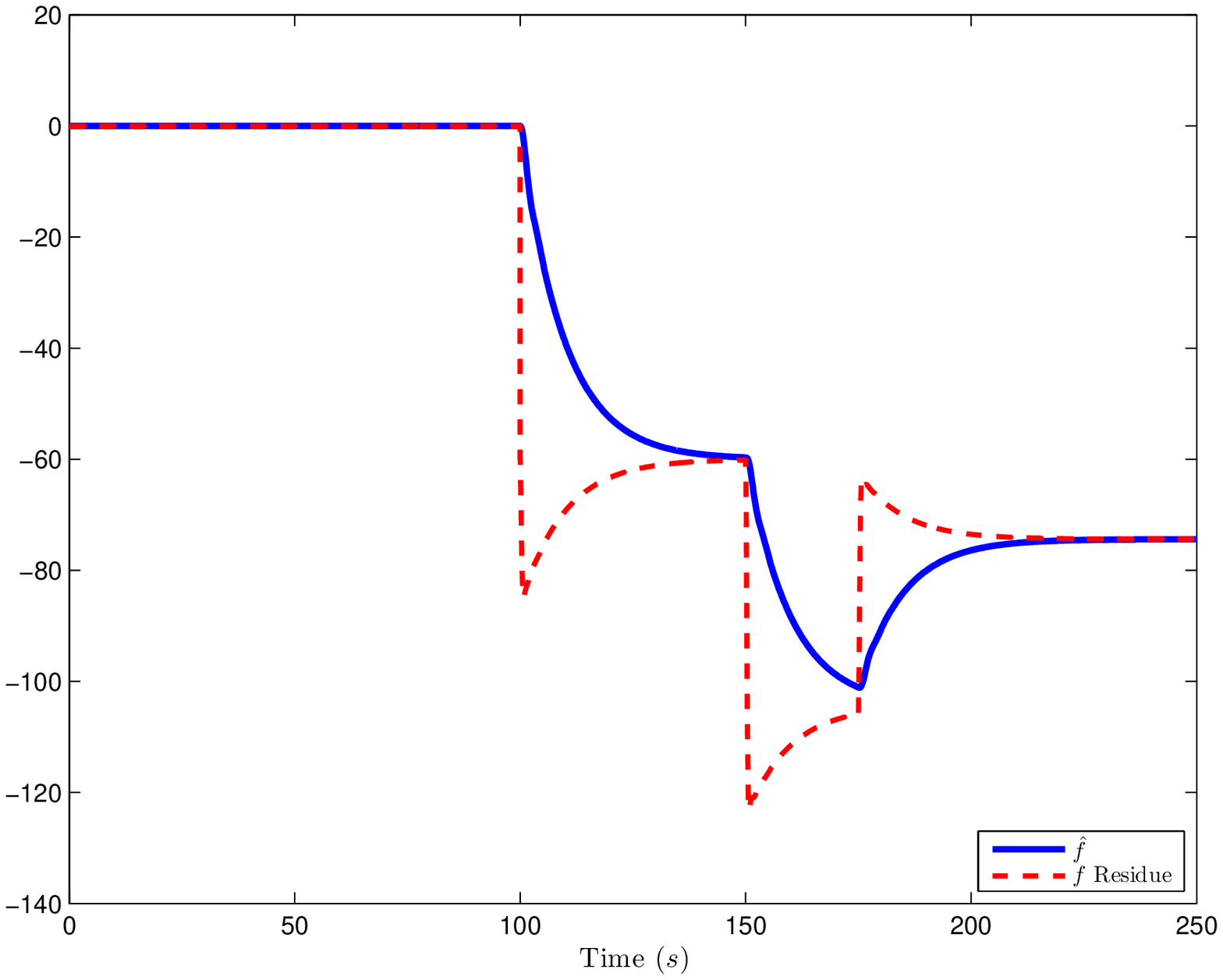}
		\caption{}
		\label{Fig 6b}
	\end{subfigure}
	\\
    \begin{subfigure}[t]{0.425\linewidth}
		\centering
		\includegraphics[width=\textwidth]{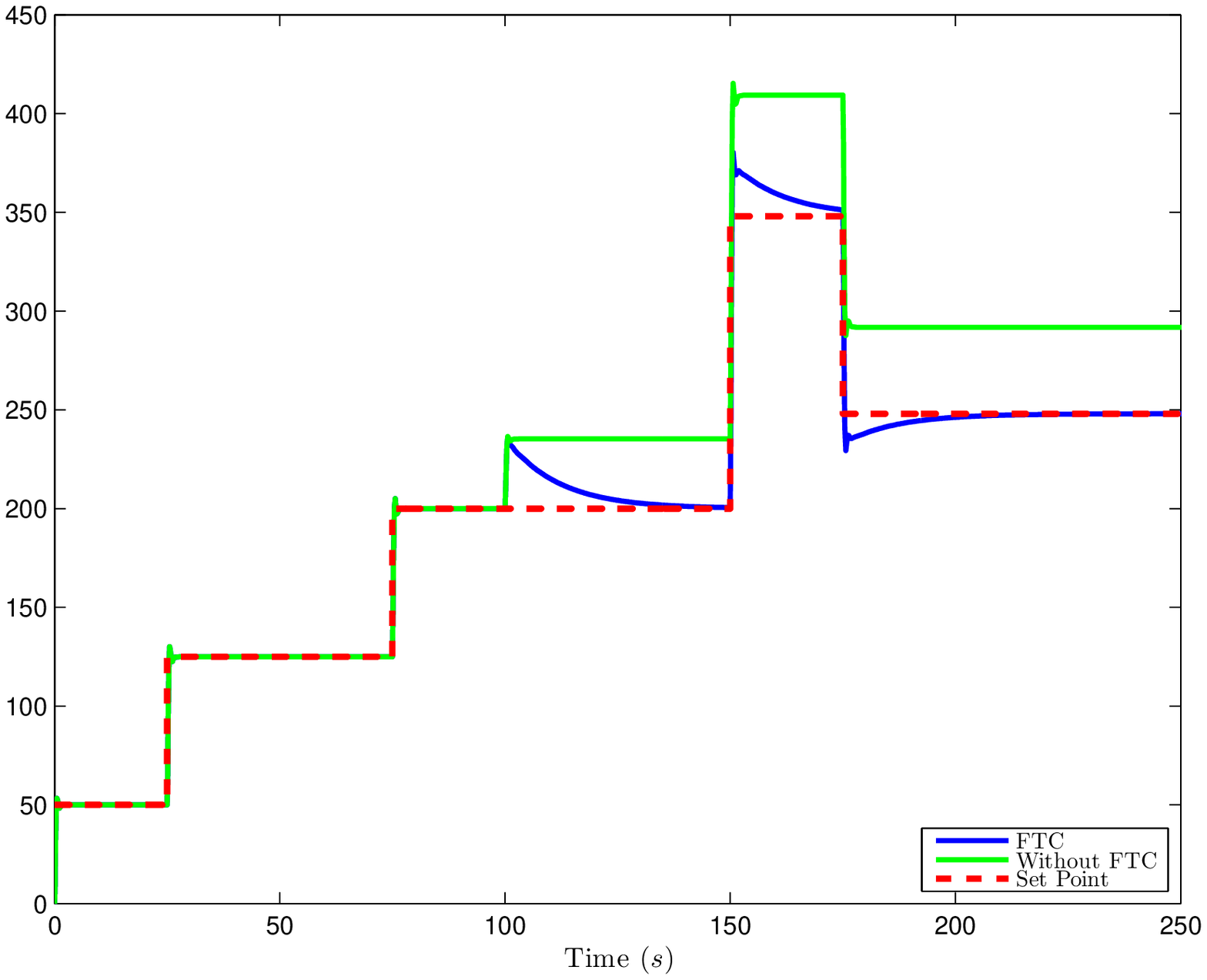}
		\caption{}
		\label{Fig 6c}
	\end{subfigure}
	~
    \begin{subfigure}[t]{0.425\linewidth}
		\centering
		\includegraphics[width=\textwidth]{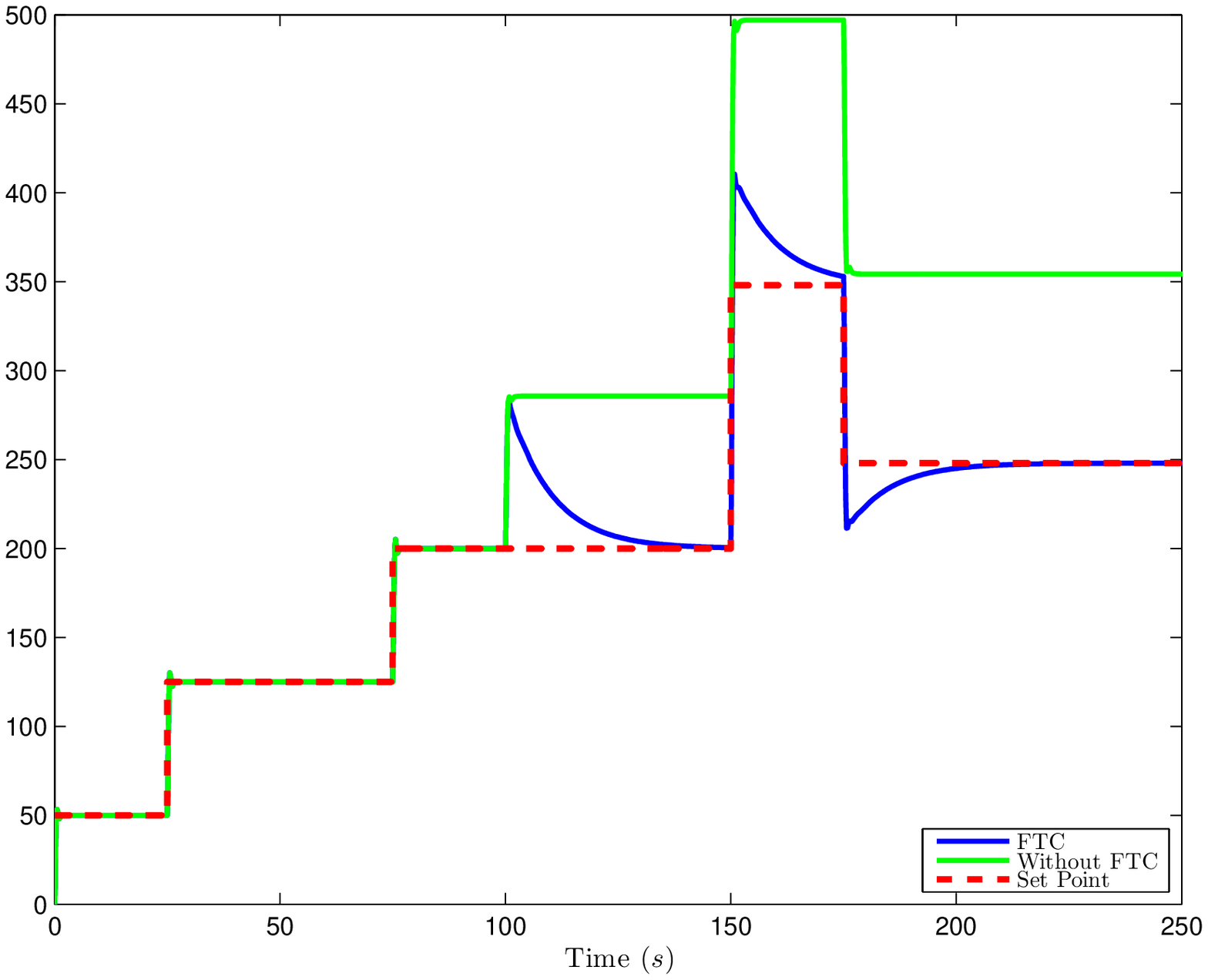}
		\caption{}
		\label{Fig 6d}
	\end{subfigure}
	\label{Figure6}
	\caption{(a). Closed-loop with PI; (b). $\hat{f}$ against $f$ residue; (c). $85\%$ sensitivity; (d). $70\%$ sensitivity}
\end{figure*}

The test of open-loop scheme depicts the rewarding results in designing the mathematical ideas approaching the true system, constituting slightly just over $1\%$ error in average as illustrated in Table. \ref{Table2}. The closed-loop scenario in Fig. (\ref{Fig 6a}) also indicates the tracking to the set-point with slight overshoot while some rising set-point comes. This implies that the gain of the controller suit the system. Furthermore, the sensing element is supposed to be failure to draw the true system with two divergent outline related to sensitivity error, comprising $85\%$ and $70\%$ in certain iteration step. Those two patterns are simulated in Fig. (\ref{Fig 6c}) and (\ref{Fig 6d}) which conclude that in the $100^{th}-$step the FTC could deal with the faulty states leading to trace the set-point whilst the system without FTC is unable to handle this situation and keeps performing the wrong tracking. The greater the faulty scenario will be, the greater the fault is created by the only-controller system. The ability of balancing the system constitutes FTC the method which is robust and preferable in industrial system if the faults on either sensor or actuator, even both, are occurred either in different time or simultaneously. Finally, Fig. (\ref{Fig 6b}) infers the tracking from observer in terms of fault compared to the residual fault. The further research is to propose estimation methods to cope with unobservable system.

\section{Conclusion}
Mathematical designs of control system from, sensor, actuator, to scrubber have been proposed along with observer and fault-tolerant control. Closed-loop system indicates the rewarding performance to trace the set-point. The performance under FTC could deal with sensor fault in $100^{th}$ iteration. It could follow the set-point (blue) compared to that of without FTC (PI only) which keep demonstrating the faulty values with two different scenarios of faulty sensitivity. The created observer also emphasizes the rewarding results by the ability of fault estimation to cope with the $f$ residue. The following project is to place the faults on both actuator and sensor along with some variations of estimation methods apart from the conventional which could cope with the unobservability.

\fontsize{10pt}{10pt}\selectfont
\bibliographystyle{IEEEtran}
\bibliography{reference.bib}

\end{document}